\documentclass[apj]{emulateapj}

\usepackage{xspace}    %package and command to prevent missing spaces after custom newcommands
\usepackage{natbib,amsmath}
\slugcomment{Mawet et al.~2016, in preparation}

\shorttitle{Fiber Injection Unit for HDC.}
\shortauthors{Mawet et al.}

\begin{document}


\title{Observing Exoplanets with High-Dispersion Coronagraphy.\\
II. Demonstration of an Active Single-Mode Fiber Injection Unit}


\author{D.~Mawet\altaffilmark{1,2}, G.~Ruane\altaffilmark{1,3}, W.~Xuan\altaffilmark{1}, D.~Echeverri\altaffilmark{1}, N.~Klimovich\altaffilmark{1}, M.~Randolph\altaffilmark{1}, J.~Fucik\altaffilmark{1}, J.K.~Wallace\altaffilmark{2}, J.~Wang\altaffilmark{1}, G.~Vasisht\altaffilmark{2}, R. Dekany\altaffilmark{1}, B.~Mennesson\altaffilmark{2}, E.~Choquet\altaffilmark{2}, J.-R.~Delorme\altaffilmark{1} and E.~Serabyn\altaffilmark{2}}


\altaffiltext{1}{Department of Astronomy, California Institute of Technology, 1200 E. California Blvd, MC 249-17, Pasadena, CA 91125}
\altaffiltext{2}{Jet Propulsion Laboratory, California Institute of Technology, 4800 Oak Grove Drive, Pasadena, CA 91109}
\altaffiltext{3}{NSF Astronomy and Astrophysics Postdoctoral Fellow}

\email{dmawet@astro.caltech.edu}


\begin{abstract}
High-dispersion coronagraphy (HDC) optimally combines high contrast imaging techniques such as adaptive optics/wavefront control plus coronagraphy to high spectral resolution spectroscopy. HDC is a critical pathway towards fully characterizing exoplanet atmospheres across a broad range of masses from giant gaseous planets down to Earth-like planets. In addition to determining the molecular composition of exoplanet atmospheres, HDC also enables Doppler mapping of atmosphere inhomogeneities (temperature, clouds, wind), as well as precise measurements of exoplanet rotational velocities.
%high-dispersion coronagraphy (HDC) is on the critical path to the full characterization of exoplanet atmospheres across a broad range of masses from giant gaseous planets down to Earth-like planets. Besides molecular composition, HDC also enables Doppler mapping of atmosphere inhomogeneities (temperature, wind) and surface features, as well as precise measurements of rotational velocity. 
Here, we demonstrate an innovative concept for injecting the directly-imaged planet light into a single-mode fiber, linking a high-contrast adaptively-corrected coronagraph to a high-resolution spectrograph (diffraction-limited or not). Our laboratory demonstration includes three key milestones: close-to-theoretical injection efficiency, accurate pointing and tracking, on-fiber coherent modulation and speckle nulling of spurious starlight signal coupling into the fiber. Using the extreme modal selectivity of single-mode fibers, we also demonstrated speckle suppression gains that outperform conventional image-based speckle nulling by at least two orders of magnitude. 
\end{abstract}





\section{Introduction}\label{sec:intro}
At the crossroads between planetary science and astronomy, the field of exoplanet studies is undergoing unprecedented growth. Aided by numerous dedicated ground-based and space-based facilities and instruments, thousands of new worlds have been discovered over the past two decades. The vast majority of detections so far have been through indirect measurements that take advantage of the gravitational influence of planets on their host star, that of other stars on the space-time continuum, or simply the photometric dimming of starlight as the planet eclipses our line of sight. The techniques exploiting these effects, namely Doppler radial velocimetry, micro-lensing, and transit photometry, are now routinely employed for exoplanet detection and have ushered in a new era in planetary science called exoplanetology. Exoplanetology has put the Solar System into a universal perspective, and finally provides an opportunity to understand planet formation and evolution in statistical terms. 

Direct detection has eluded the exoplanet community for many years, mainly due to the stark difficulty associated with disentangling the signal of an exoplanet from its host star. The requirements to directly image a planet stretch the limits of current facilities and instruments in all possible directions: angular resolution, sensitivity, dynamic range, precision, and stability. The advent of large ground-based and space-based telescopes, adaptive optics, new infrared and optical detector technologies, and modern computing have admittedly done little to overcome these challenges. A niche technology borrowed from solar astronomy, namely coronagraphy, once held the promise of revolutionizing the field, but the long-awaited breakthrough is slow to unfold. 

Coronagraphy was invented in the 1930s by French astronomer Bernard Lyot \citep{Lyot1939} to observe and characterize the solar corona without the need for natural eclipses. The principle of coronagraphy is simple and aims, by way of a device blocking the glare of the Sun, at reducing the contrast of the scene to be within the dynamic range of the detectors. Coronagraphs now come as standard equipment on any high-contrast imaging instrument, paired with wavefront control systems (adaptive optics), including deformable mirrors controlled in closed loop via a series of dedicated wavefront sensors. Downstream from the high-contrast  equipment are classical imaging cameras, and/or low spectral resolution integral field spectrographs (IFS). 

A key strategy to differentiate between planets and leftover speckles of residual starlight is to modulate the planet signal against the background of dynamic and quasi-static speckles. Many differential imaging techniques have been devised to mitigate speckle noise, such as: angular differential imaging (ADI), spectral/simultaneous differential imaging (SDI), dual-band differential imaging (DBI), reference star differential imaging (RDI), polarization differential imaging (PDI), coherent differential imaging (CDI), orbital differential imaging (ODI) and binary differential imaging (BDI). ADI and SDI are by far the most successful, but present significant challenges at very small inner working angles, owing to signal self-subtraction effects \citep{Mawet2012}.

\begin{figure*}[t!]
  \centering
\includegraphics[width=17cm]{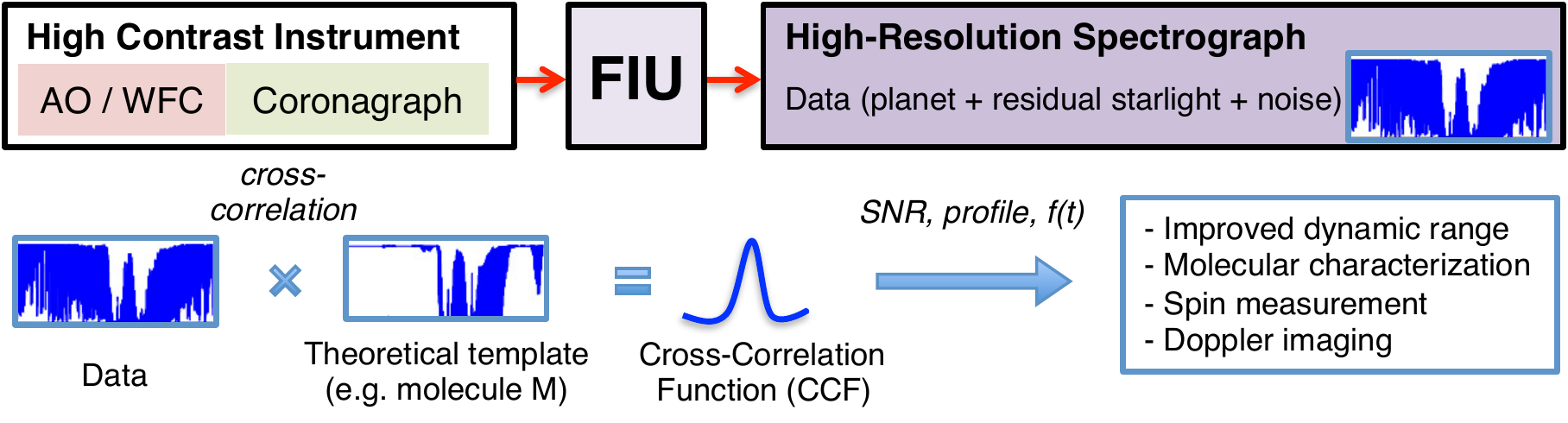}
  \caption{High-dispersion coronagraphy (HDC) concept. A classical high-contrast instrument, with an adaptive optics (AO) or wavefront control (WFC) system followed by a starlight suppressing coronagraph, is linked to a high-resolution spectrograph uby a fiber injection unit (FIU). The data processing steps are as follows: the raw data (planet+residual starlight unfiltered by the high-contrast instrument, and various noise contributors such as photon shot noise, readout noise, background noise, etc.) is cross-correlated with a theoretical template yielding a new observable, called here the cross-correlation function (CCF). The CCF profile provides improved dynamic range for detection and molecular characterization. Its broadening with respect to the instrument line profile is a direct measure of the planet's spin rotation. The variation of the line profile morphology over the rotation period enables Doppler imaging. \label{fig:concept}}
\end{figure*}

Here we propose and demonstrate a new concept that optimally combines high-contrast imaging techniques and high-resolution spectroscopy, called for the sake of simplicity high-dispersion coronagraphy (HDC). The promise of HDC is the cumulative gain in the performance offered by each technique, as first suggested by \citet{Riaud2007} and more recently refined by \citet{Snellen2015}. The reason being that high-resolution spectroscopy sidesteps the problem of speckle noise, since speckle noise has a low spectral resolution signature \citep{Krist2008} and is effectively part of the continuum at high spectral resolution. Moreover, the planet signal will be shifted in frequency (velocity) space with respect to the star signal due to the Doppler effect induced by the orbital motion of the planet around its host star, enabling spectral lines to be disentangled from one another. Thus, HDC is perhaps the only differential method that will approach the photon noise limit.

In this paper, we present a new concept for feeding a filtered beam of planet light to a high-resolution spectrograph (Fig.~\ref{fig:concept}). The framework of this proof-of-concept is the Keck Planet Imager and Characterizer project (KPIC), a planned upgrade to the W.M. Keck Observatory adaptive optics system and high-contrast instrument suite \citep{Mawet2016}. KPIC will serve as a pathfinder for future high-contrast spectroscopic instruments for large ground- and space-based facilities: the Thirty Meter Telescope (TMT), the European-Extremely Large Telescope (E-ELT), the Giant Magellan Telescope (GMT), NASA's Habitability Explorer (HabEx), and the Large UV Optical InfraRed (LUVOIR) telescopes. 

\section{High-contrast high-resolution spectroscopy of exoplanets}\label{sec:HCHRS}

Now that thousands of exoplanets have been discovered, detailed characterization of these planets is the logical next step. The leading detection methods based on radial velocities (RV) and transits provide only the mass and/or size of the planet. With these measurements, bulk density and chemical composition may be inferred with exoplanet internal structure models. However, this approach suffers from degeneracies, highlighting the need for directly measuring their chemical compositions. 

Detailed diagnoses of the chemical composition of exoplanet atmospheres \citep[see e.g.][]{Barman2011, Konopacky2013, Barman2015} remain a challenge because of the small angular separation and high contrast between exoplanets and their host stars. Both constraints are mitigated by a high-contrast imaging system, which usually consists of an extreme adaptive optics (AO) system and a coronagraph. Current state-of-the-art high-contrast imaging systems such as the Gemini Planet Imager at the Gemini South telescope \citep{Macintosh2015} and SPHERE at the Very Large Telescope \citep{Beuzit2008} are able to achieve $10^{-3}$ to $10^{-4}$ raw starlight suppression levels at a few tenths of an arcsecond, allowing detections and very low-resolution spectroscopy (spectral resolution R$\simeq 50$) of gas giant planets and brown dwarfs orbiting nearby young stars.

\citet{Riaud2007} and \citet{Snellen2015} suggested that contrast sensitivity may be further improved by coupling a high-dispersion spectrograph with a high-contrast imaging system. In this scheme, the high-contrast imaging system serves as a spatial filter to separate the light from the star and the planet, and the high-dispersion spectrograph serves as a spectral filter taking advantage of differences between the stellar and planetary spectra, including absorption lines and radial velocities (see Fig.~\ref{fig:concept}). 

Using high-dispersion spectroscopy as a way to spectrally isolate the planet signal has been successfully demonstrated by a number of integrated light studies. Indeed, high-resolution transmission spectroscopy has been used to detect molecular gas in the atmosphere of transiting planets \citep{Snellen2010, Birkby2013, deKok2013}. At a high spectral resolution, resolved molecular lines may be used to study day-to-night side wind velocity \citep{Snellen2010} and verify 3D exoplanet atmospheric circulation models \citep{Kempton2014}. The spectral lines of a planet may also be separated from stellar lines with sufficient differences in radial velocities ($>50$ km/s), breaking the degeneracy between the true planet mass and orbital inclination \citep{Brogi2012, Brogi2013, Brogi2014, Lockwood2014}. Moreover, high-resolution spectroscopy has led to the first measurement of planet rotational velocity \citep{Snellen2014}. While not yet feasible on exoplanets yet, high-resolution spectroscopy has helped generate the first global cloud map of brown dwarf Luhman 16 B via the Doppler imaging technique \citep{Crossfield2014}.

High-resolution spectroscopy is poised to become even more powerful when combined with high-contrast imaging. The signal-to-noise ratio (SNR) achieved by an HDC instrument is to first order \citep{Snellen2015}:
\begin{equation}\label{eq1}
S/N = \frac{\eta S_\mathrm{planet}} {\sqrt{S_\mathrm{star}/K + \sigma^2_\mathrm{bg} + \sigma^2_\mathrm{rn} + \sigma^2_\mathrm{dark}}} \sqrt{N_\mathrm{lines}},
\end{equation}
where $S_\mathrm{planet}$ is the planet signal making it to the spectrograph with efficiency $\eta$, $S_\mathrm{star}$ is the signal from the star (both in units of photo-electrons per pixel), $K$ is the suppression factor of the star at the planet's position, and $\sigma^2_\mathrm{bg}$, $\sigma^2_\mathrm{rn}$, and $\sigma^2_\mathrm{dark}$ are the photon shot noise from the sky and telescope background, the readout noise, and the dark current noise, respectively. $\sqrt{N_\mathrm{lines}}$ is a multiplication factor that takes into account the number and strength of the individual planet lines targeted, which is a defining strength of high-resolution spectroscopy \citep{Snellen2015}.

The planet/star contrast sensitivity achieved by integrated-light high-dispersion spectroscopy is currently demonstrated at the $10^{-4}$ level, which corresponds to the stellar photon noise limit \citep{Snellen2015}. When coupled with a state-of-the-art high-contrast imaging system with a raw starlight suppression of $10^{-3}$ or better, a high-contrast high-dispersion spectroscopy instrument can potentially exceed $10^{-7}$ planet/star contrast, providing superior sensitivity than a high-contrast imaging system or a high-dispersion spectrograph alone. This would allow the physical and chemical processes taking place on an exoplanet to be studied in unprecedented details. 

It is important to note that high-spectral-resolution observations of a single spatial resolution element render spatial speckle variations (spatial speckle noise) irrelevant. Since the spectral signature of stellar speckles is a very smooth and a slowly varying function of wavelength, it becomes part of the continuum at very high spectral resolutions \citep{Krist2008}. Thus, the dominant limiting factor in low-resolution high-contrast imaging, spatial speckle noise, is obviated by HDC.

Ground-based HDC observations will enable the detection of multiple molecular species and their resolved spectral lines in the J, H, K, L, and M bands (Wang et al.~2017, submitted). Currently known directly-imaged exoplanets (e.g., HR 8799bcde, 51 Eri b, ROXs 42B b, ROXs 12 b, $\beta$ Pictoris b) will be prime targets for HDC observations. Together with observations from JWST in the near-to-mid infrared wavelengths (at much larger wavelengths, e.g., $> 5 \mu$m), ground-based HDC observations will yield abundances \citep{Brogi2016}, remove the degeneracy of temperature and pressure profiles and thus provide more details on the presence and formation of cloud/haze, which is a critical step forward in understanding the physical and chemical processes in exoplanet atmospheres.

\begin{figure*}[!t]
  \centering
\includegraphics[width=16cm]{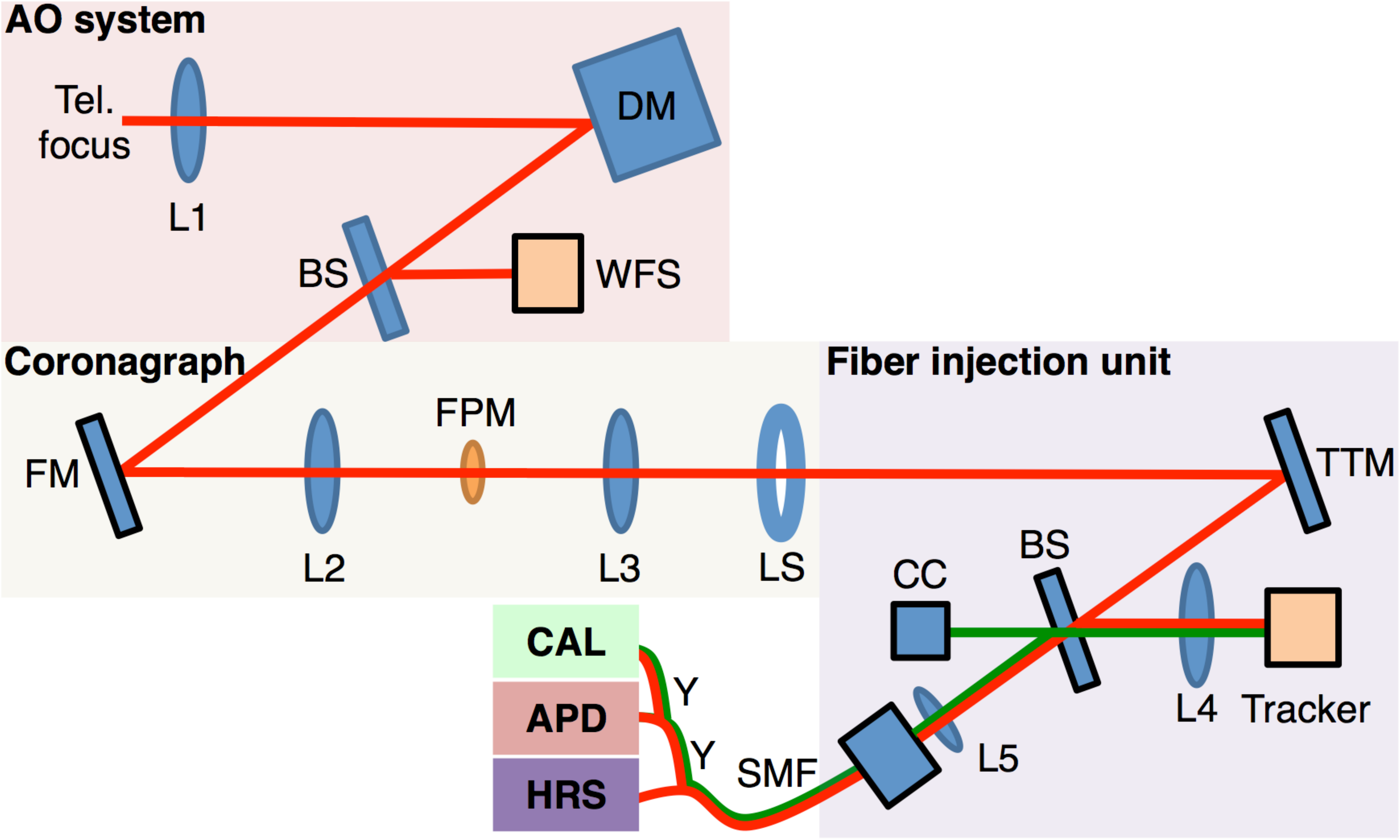}
  \caption{Layout of a typical high dispersion coronagraph (HDC) and our laboratory setup, consisting of an AO system with a deformable mirror (DM) and wavefront sensor (WFS), a coronagraph with a focal plane mask (FPM) and Lyot stop (LS), and a fiber injection unit with a tip-tilt mirror (TTM), beamsplitter (BS) or dichroic, tracking camera, corner cube (CC), single-mode fibers (SMF), and optics to image the scene on the camera and inject planet light into the fiber(s). The fibers feed light to the high-resolution spectrograph (HRS) and an avalanche photodiode (APD). A calibration source (CAL) is used to back propagate light through the SMF for tracking purposes. \label{fig:layout}}
\end{figure*}

\section{Fiber injection unit concept}\label{sec:FIU_concept}
Here we propose to link the coronagraph instrument and the high-resolution spectrograph with a fiber injection unit (FIU), illustrated in Fig.~\ref{fig:layout}. The purpose of the FIU is to couple planet light into a single mode fiber (SMF) and maintain accurate alignment throughout long-exposure observations (up to several hours). The pointing accuracy and stability is achieved through active sensing and control of the planet and fiber positions using a scheme similar to \citet{Colavita1999}.

An actuated tip-tilt mirror (TTM) is used to align the planet image position with the tip of the SMF, whose relative locations are determined by simultaneously imaging the scene and the SMF on to a tracking camera. A beamsplitter (BS) or dichroic reflects part of the science beam to the tracking camera directly after the TTM. To locate the SMF, a calibration source (CAL) is retro-fed through the fiber by means of optical circulators or Y-couplers. The BS reflects light from the SMF towards a corner cube (CC) retroflector, which sends the beam back through the BS and towards the tracking camera. A beacon image is formed on the tracking camera at the location of the SMF. The beacon is used to determine the TTM settings to co-align the object image and the SMF. Alternatively, the CAL source may feed a separate SMF creating a beacon nearby the spectrograph fiber tip with well calibrated relative positions. 

The FIU is also designed to provide feedback mechanisms for starlight suppression using the upstream deformable mirror (DM). An optional low-noise single pixel detector (e.g. an avalanche photodiode; APD) may be used to monitor the starlight leaking into the fiber at high-speed ($>$10 kHz) and drive a control loop that minimizes leaked starlight in real time. We have demonstrated both the optical alignment procedure and real time wavefront control concepts in the laboratory. 

%The FIU optical system includes a tip-tilt mirror, a beamsplitter (or dichroic), a tracking camera, a corner cube, a set of single-mode fibers (SMF), and optics to image the scene on the camera and inject planet light into the fiber(s). The FIU feeds a high-resolution spectrograph and an optional single-pixel detector (e.g., avalanche photodiode) used for single-mode wavefront control.   A tunable fraction of the object's light is reflected off the beamsplitter (or dichroic) to the camera. A calibration light is retro-fed into the single mode fiber, propagates backward, gets launched and collimated in the injector, and reflects off the other surface of the same beamsplitter into a corner cube (retroreflector). The corner cube reflects the beam back directly towards the beamsplitter, and thus the tracking camera, forming a beacon image of the fiber location. The actuated tip-tilt mirror is used to co-align the object image with the fiber beacon.

\section{Laboratory setup}\label{sec:lab_setup}
\begin{figure}[!t]
  \centering
\includegraphics[width=8.5cm]{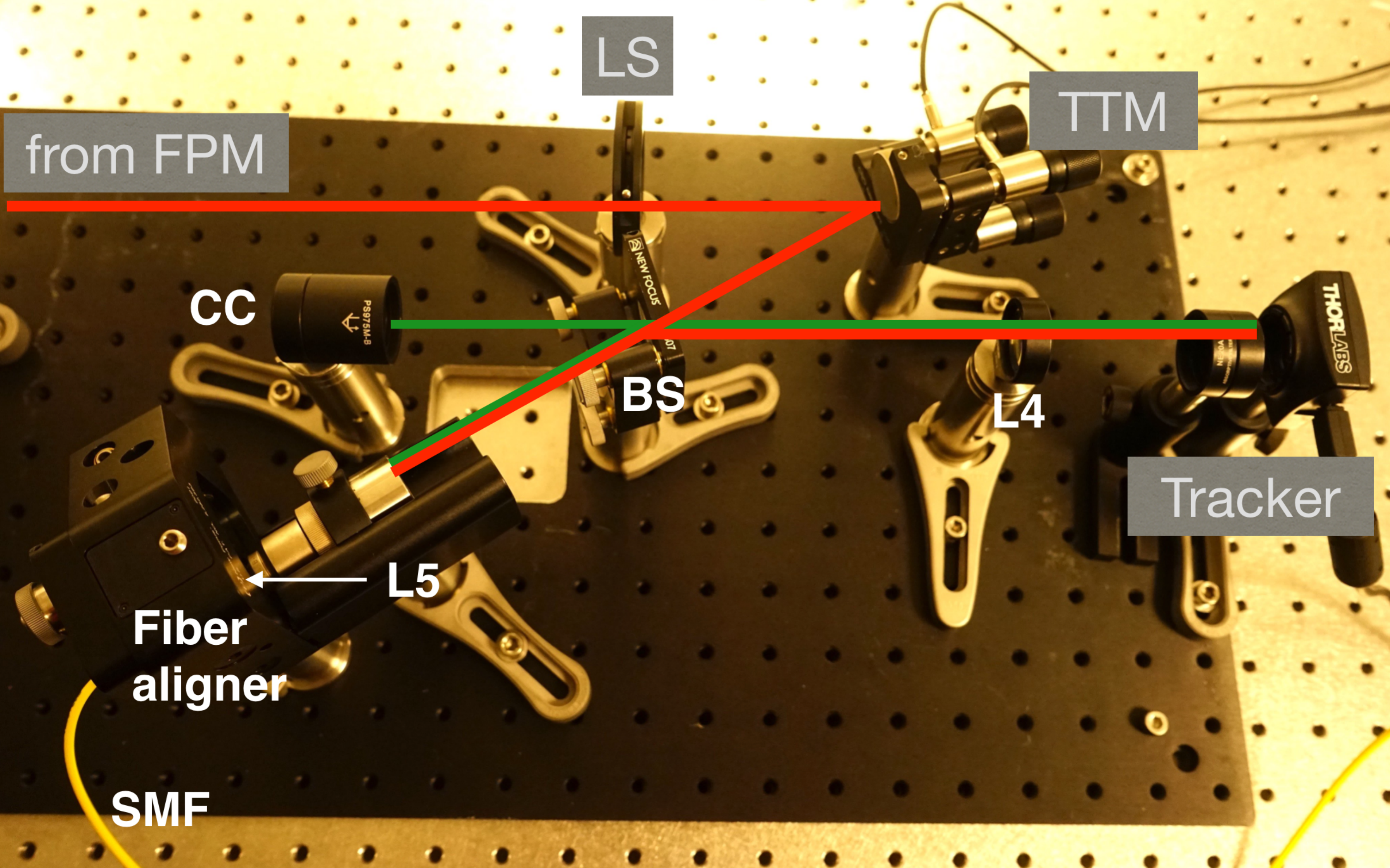}
  \caption{Top view of the FIU prototype on Caltech's High Contrast Spectroscopy Testbed for Segmented telescopes (HCST). \label{fig:lab_pic}}
\end{figure}

Our laboratory setup consists of a telescope simulator, followed by an adaptive optics (AO) system, a coronagraph, and the FIU prototype (see Figs.~\ref{fig:layout} and \ref{fig:lab_pic}). The telescope simulator images simulated star and off-axis planetary sources, generated by a Thorlabs 635 nm laser diode and a filtered NKT Photonics supercontinuum white light source (narrowband filter centered at 650 nm), respectively. The adaptive optics system is made up of a Boston Micromachines 144-actuator MEMS DM, pellicle BS, and a Shack-Hartmann wavefront sensor (Thorlabs AOK1-UM01). 

%To simulate the starlight we used a 635 nm laser diode from Thorlabs. To simulate the incoherent planet light we used an NKT Photonics supercontinuum white light source, filtered by a narrowband filter centered at 650 nm. The planet light is launched into the beam in collimated space.

%Downstream from the telescope and source simulator, our current setup includes an adaptive optics system, consisting of the Thorlabs AO kit (AOK1-UM01). The Thorlabs AO Kit is a complete adaptive optics imaging solution, including a 144-actuator MEMS deformable mirror from Boston Micromachine Inc., a Shack-Hartmann wavefront sensor, control software, and optomechanics for assembly. 

The AO system is followed by a classical 3-plane coronagraph with a vortex focal plane mask (FPM). The vortex coronagraph is a phase-based coronagraph enabling high-contrast imaging at small angular separations, while conserving high off-axis throughput \citep{Mawet2005}. The vortex coronagraph is currently in operations at Palomar \citep{Mawet2010,Serabyn2010,Mawet2011,Bottom2015,Bottom2016a}, VLT \citep{Mawet2013}, Subaru, Keck \citep{Absil2016,Serabyn2017,Mawet2017}, and Large Binocular telescopes \citep{Defrere2014}. The particular vortex mask used here has a topological charge of 4, which applies a phase ramp of the form $e^{i 4\theta}$, where $\theta$ is the azimuthal angle in the focal plane. The effective inner working angle (i.e. the angle for 50\% off-axis transmission) of the charge 4 vortex coronagraph is $\sim1.7\lambda/D$, where $\lambda$ is the central wavelength, and $D$ is the telescope diameter.

Downstream from the coronagraph's Lyot stop lies the fiber injection unit described in Sect.~\ref{sec:FIU_concept}. We use a three-axis tip-tilt mirror from Newport, actuated by computer controlled Thorlabs piezoactuators. The FIU BS is a 50\%-50\% beamsplitter (at the telescope, we plan to use more optimal splitting ratios and/or dichroic beamsplitters). The tracking camera is a CMOS sensor from Thorlabs. The single-mode fiber is mounted on a Newport Post-Mount Singlemode Fiber Aligner. The corner cube and other optical elements are off-the-shelf Newport and Thorlabs products. For the purposes of this demonstration, we used a Newport Si photodiode power meter in lieu of the high-resolution spectrograph. The back-end calibration source is a fiber-coupled 635 nm~laser diode.

\begin{figure*}[!t]
  \centering
\includegraphics[width=6cm]{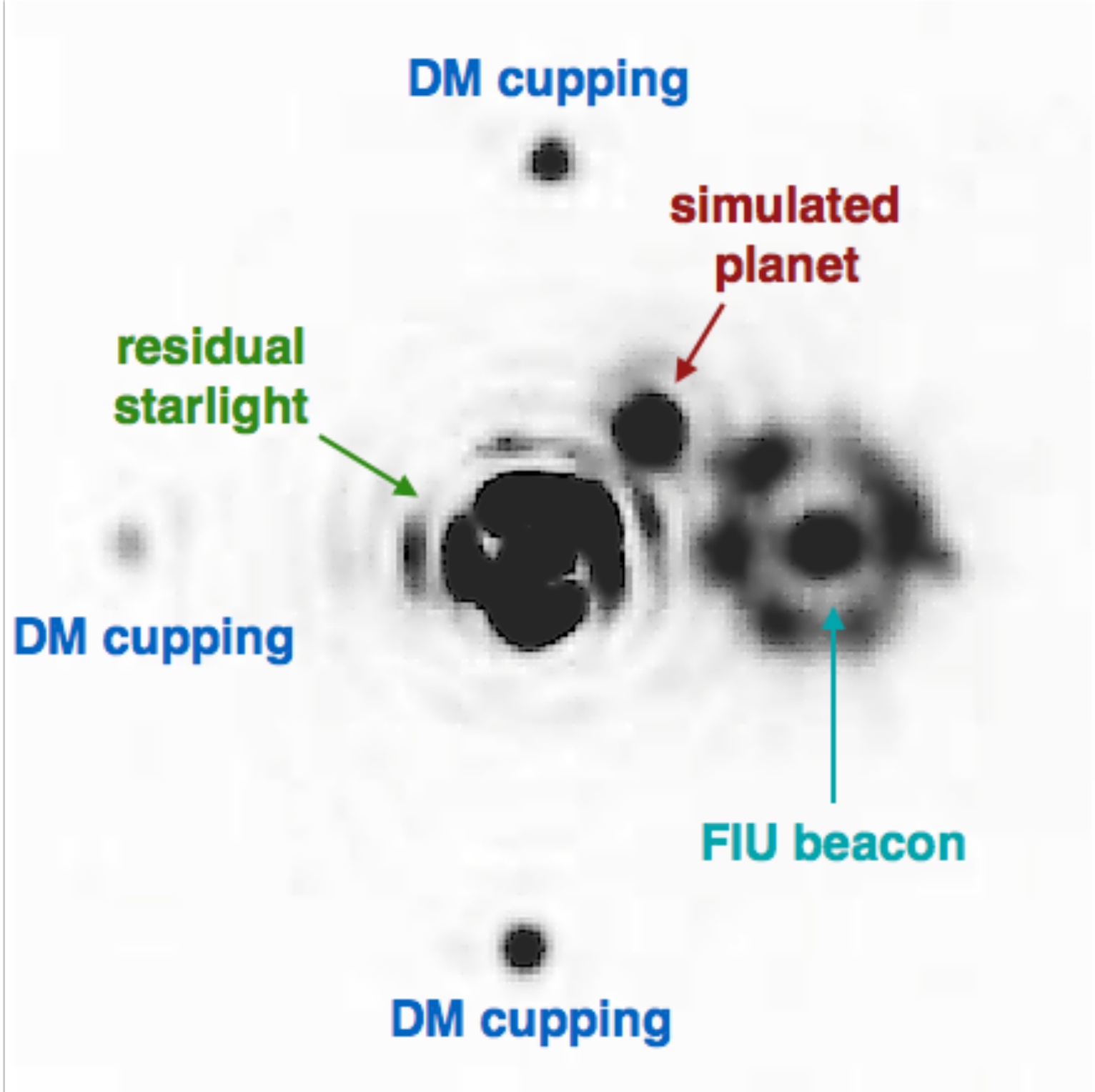}
\includegraphics[width=11cm]{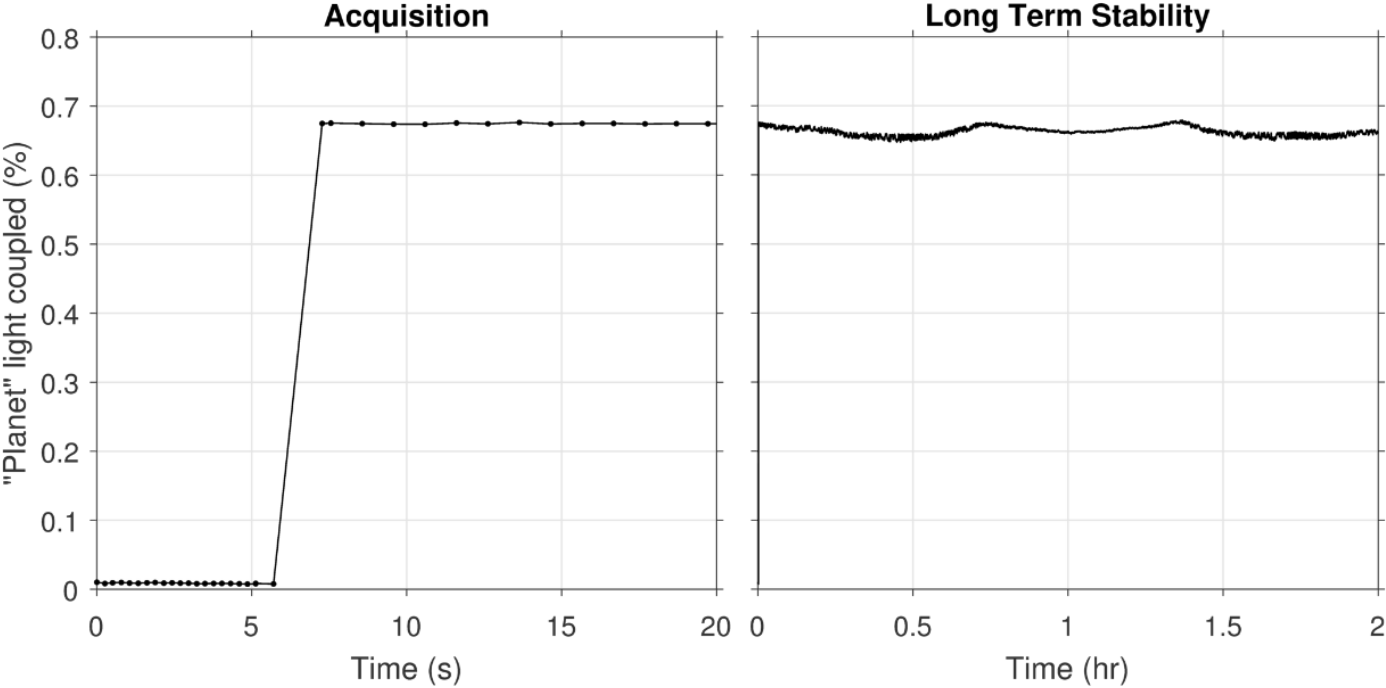}
  \caption{Left: Image from the tracking camera, showing residual starlight after the charge-4 vortex coronagraph, along with our simulated off-axis planet and the beacon calibration lamp. Middle: Acquisition sequence showing the quick injection of the off-axis planet signal into the SMF after determining the position of the SMF via the FIU beacon. Right: Injection stability over two hours. \label{fig:coupling}}
\end{figure*}

\section{Results}\label{sec:results}

To validate our new FIU concept, we conducted a series of experiments. We first demonstrated the coupling of starlight and planet light into the fiber using manual spiral scans. Then, the co-alignment procedure using the beacon image was demonstrated by injecting planet light into the fiber in a reproducible manner. We then validated the wavefront control procedure to minimize the amount of starlight coupling into the fiber along with the planet light using a technique akin to speckle nulling \citep{Borde2006,Bottom2016b}.

\subsection{Planet injection}\label{sec:Planet_injection}

Our first demonstration consisted of injecting a point source image into the fiber with an efficiency close to the theoretical limit. A single mode fiber's fundamental mode is nearly a Gaussian \citep{Shaklan1988}, while our optical system generates an Airy function of the form $J_1(r)/r$, which is the result of the Fourier transform of a circular unobscured input pupil, and thus the point spread function (PSF) of our optical setup. The monochromatic injection efficiency $\eta$ is the modulus squared of the overlap integral between the incident electric field $A(x,y)$ and the fundamental mode of the SMF $\mathrm{HE}_{11}$: 
\begin{equation}\label{eq2}
\eta = \frac{\left| \iint \mathrm{HE}_{11}(x,y) A^*(x,y) dx dy\right|^2}{\iint |\mathrm{HE}_{11}(x,y)|^2 dx dy \iint |A(x,y)|^2 dx dy}.
\end{equation}
% \begin{equation}\label{eq2}
% \eta = \frac{\left| \iint \mathrm{HE}_{11}(x,y) A^*(x,y) dx dy\right|^2}{\iint \mathrm{HE}_{11}(x,y)\mathrm{HE}^*_{11}(x,y) dx dy \iint A(x,y)A^*(x,y) dx dy},
% \end{equation}}

The theoretical maximum injection efficiency is $81\%$ for an ideal circular, unobstructed pupil \citep{Shaklan1988}. Due to aberrations in the system, the impinging field is not exactly an Airy function, and so the overlap integral inevitably yields lower injection efficiencies \citep{Wagner1982,Toyoshima2006}. We note that the theoretical maximum injection efficiency may be increased by apodizing the impinging beam into a Gaussian function \citep{Jovanovic2015}. 

%The injection process for the system is routine and straightforward. We would insert the SMF into the calibration laser while looking at the beacon and the planet in the CCD image. Fig.\ref{fig3} (left) shows at the center the residual starlight filtering through the charge 4 vortex coronagraph, along with the simulated planet light and the beacon retro-fed from the laser metrology.

Figure \ref{fig:coupling} (left) shows the image on the tracking camera with the residual starlight concentrated at the center of the image as well as the simulated planet and FIU beacon. The computer control of the tip-tilt mirror yielded consistent and reproducible injection efficiencies between 65\% and 70\%, which was deemed sufficient for demonstration purposes (Fig.~\ref{fig:coupling}, middle). The difference between our measured efficiencies and the theoretical limit can be traced to optical aberrations and Fresnel losses at air/glass interfaces in the injector and on the tip of the fiber. The planet injection is very stable over hour timescales as shown in Fig.~\ref{fig:coupling} (right).

On-sky demonstrations have so far yielded consistent and reliable results, but nowhere near the theoretical limit due to residual wavefront error after adaptive optics correction \citep{Jovanovic2016, Bechter2016}. The coupling efficiency results achieved in the lab therefore represents an upper limit on what can be achieved on sky. Using Subaru/SCExAO, the state-of-the-art on-sky injection efficiency was reported by \citet{Jovanovic2016} to be $\simeq 50\%$ in the H band.

%With manual drift searches of the tip-tilt mirror, we put the planet at the position of the beacon, making sure that their respective centroids roughly overlap. This would usually produce half or more of our maximum injection efficiency, so a second search with the power meter is performed, where we disconnect from the reference laser, insert the fiber into the Si photodiode, and attempt to reach maximum power by moving in small steps around the centroid. The above steps could be done iteratively to ensure that the correct signal is injected, e.g., the planet rather than an adjacent speckle.

%Our first test using manual scans of the tip-tilt mirror yielded consistent and reproducible injection efficiencies between 60\% and 70\%, which was deemed sufficient for demonstration purposes (Fig.~\ref{fig3}, middle). Injection stability is very good over hour timescales as shown in Fig.~\ref{fig3} (right).

\subsection{Starlight rejection: speckle nulling}
A necessity in imaging--let alone characterizing--exoplanets is to suppress residual starlight (speckles) in the final image plane as much as possible. Speckles are caused by optical aberrations incurred as light travels through Earth's atmosphere and imperfect optics in the imaging system. Speckles are the nemesis of exoplanet imaging, since they might appear similar to or overwhelm the planet signal, precluding both discovery and characterization. 

In short, speckle nulling is the process of destructively interfering an intentionally generated anti-speckle with an existing speckle. We recall that by virtue of the Fourier transform relationship between the pupil plane and the image plane, a sinusoid with amplitude $h_0<<\lambda$ in the pupil plane will translate into a pair of conjugated anti-speckles in the image plane \citep{Malbet1995}. Specifically, we apply a cosine pattern to the DM surface with height
\begin{equation}\label{eq3}
    h = \frac{h_0}{4}\left[\cos\left(2\pi \vec{\xi} \cdot \vec{r} + \alpha \right)\right],
\end{equation}
where $h_0/2$ is the maximum surface height ($h_0$ wavefront), $\vec{\xi}$ is the spatial frequency vector and the speckle location in the image plane, $\vec{r}$ is the position vector in the pupil plane, and $\alpha$ is a constant phase offset. The intensity, position, and phase of the speckle are controlled by $h_0$, $\vec{\xi}$, and $\alpha$, respectively. It is worth noting that the speckle intensity is $(\pi h_0/\lambda)^2$ \citep{Malbet1995}. The limited number of actuators illuminated on the DM (roughly $11\times11$) in theory limits the spatial frequency to $\sim$5 cycles per pupil diameter and, therefore, the range of angular separations at which anti-speckles may be produced {$\sim$1-5~$\lambda/D$}. However, it is possible to use high-spatial frequency DM surface features, such as print-through and cupping effects, to increase this range taking advantage of harmonics (i.e. clone PSFs) that appear at integer multiples of 11~$\lambda/D$ \citep{Thomas2015}. 

%$q$ is the angle of sinusoid pattern, starting from zero at the positive x-axis, $x_0$ is the number of DM actuators per cycle, and $\alpha$ is the phase delay. In this equation, $h_0$ controls the intensity of the anti-speckle, $\alpha$ determines its phase delay, $q$ determines its distance from the star. Our equation includes extra factors of 0.5 to take into account the fact that our DM actuators only deflect in one direction, so that a zero displacement always represents the trough of the sinusoid.

%With the ability to generate anti-speckles from ~ $1\lambda/D$ - $5\lambda/D$ (limited by the number of actuators illuminated in our DM, roughly 11-by-11) away from the star, we wrote code to loop through amplitude and phase space while also varying q (angle) and x\textsubscript{0} (distance). We save images taken by the tracking CCD after a given sinusoid is applied. A reference image taken with a flat DM is used to calculate the suppression ratio (intensity with flat DM / intensity with suppressing sinusoid on DM). 

To test the viability of this method, we first performed linear searches in $h_0$, $\vec{\xi}$, and $\alpha$ to find the parameters that optimize the starlight suppression ratio as measured on the tracking camera at various speckles at $\sim2\lambda/D$, $\sim3\lambda/D$, and $\sim4\lambda/D$ from the star. The optimization of $\vec{\xi}$ is used as a fine tuning and might be superfluous since the fiber location can be known precisely. Initial searches were done in coarse increments over the full four-dimensional parameter space, but clearly show that a global minimum null exists for each speckle. Complete searches, however, take a implausibly long time (20-40 min), which would waste valuable (and expensive) telescope time. Hence, we developed an expedited optimization code that reaches the minimum in $\sim$3-5 min. This approach relies on a pattern search optimization in phase and amplitude space, after a calibration is done to generate the speckle near the optimal correct location \citep{Bottom2016b}. However, using the tracking camera images to probe suppression is not our final goal. Rather, we wish to maximize the signal-to-noise ratio of the planet signal in the spectrograph, which depends on the amount of residual starlight that is injected into the SMF.

\begin{figure*}[!t]
  \centering
\includegraphics[width=18cm]{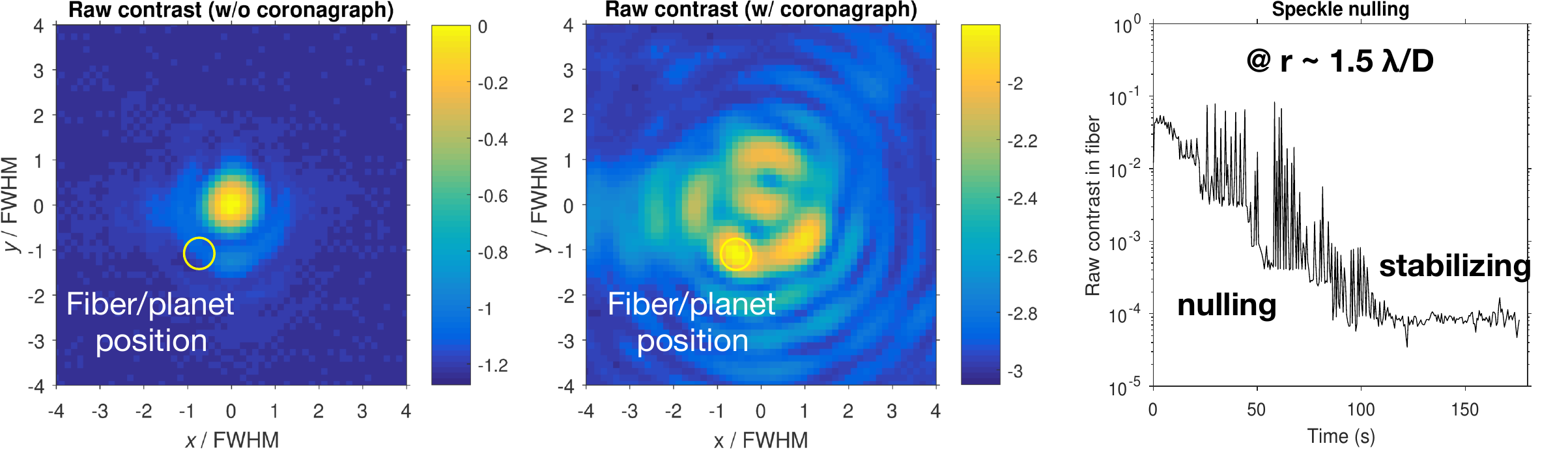}
  \caption{Experimental results in monochromatic light. Left: Stellar PSF with the coronagraph focal plane mask removed. Middle: Residual starlight with the focal plane mask aligned. The fiber and planet position is indicated in both images. The simulated planet is about 1000 times fainter than the peak starlight, and is therefore invisible in these images. Right: Nulling sequence, showing the starlight level as a function of time, while the speckle nulling loop is active. Sharp peaks in power appear in the nulling sequence as the optimization algorithm explores the four dimensional parameter space to find a deeper null. \label{fig:nulling_demo}}
\end{figure*}

\subsection{Integration: Speckle Nulling at Fiber-tip Location}
The goal of the FIU is to inject as much planet light as possible while rejecting as much starlight as possible. Unfortunately, residual aberrations in the system due to imperfect optics here, plus uncorrected atmospheric turbulence in a ground-based system, will create speckles in the image, which will also couple into the SMF and propagate into the spectrograph. Speckle noise would often overwhelm the planet signal, especially at small angular separations from the star, where speckle noise is dominant even after the spatial filtering from the coronagraph. Preventing starlight from coupling into the fiber is the best way to increase the signal-to-noise ratio, and thus efficiency of the observation.

By coupling the other end of the SMF to a photodiode, we record the total power entering the SMF after applying an optimal sinusoid pattern to the DM and calculate the suppression factor (the power of speckle with flat DM divided by the power after nulling) of a speckle at the fiber position. Specifically, we recorded monochromatic starlight suppression factors of $>1000$ with the SMF for a bright speckle located at roughly 2$\lambda/$D away from the star (see Fig. \ref{fig:nulling_demo}). We note that the measured suppression is limited by the dynamic range and noise properties of our Si photodiode.

After repeating this experiment multiple times for various speckles in the image, we find that speckle nulling with a SMF generally improves raw starlight suppression by a factor of 500-1000 beyond the nominal raw starlight suppression level produced by the wavefront control / adaptive optics system and coronagraph. The corresponding gain simultaneously measured on our tracking camera images is 5-10, which is similar to speckle nulling gains routinely demonstrated in ground-based imaging/spectroscopy systems \citep{Bottom2016b}. The presence of the planet signal at the location of the fiber and speckle has been verified not to affect nor be affected (Sect.~\ref{subsubsection:Throughput_losses_after_nulling}) by the nulling procedure. The planet signal is indeed an incoherent background and much fainter than the speckle, it does not respond to the ripple probes from the DM, and does not contribute to the sensing of speckle complex amplitude.

\subsubsection{Heuristic explanation for the suppression gain}
Standard speckle nulling techniques rely on coherent interference between the speckle and the generated anti-speckle to minimize the starlight in a particular region on the image \citep{Jovanovic2015}. As noted in Sect.~\ref{sec:Planet_injection} and Eq.~\ref{eq2}, the injection efficiency of the SMF is the modulus squared of the overlap integral between the incident E-field and the fundamental symmetric HE$_{11}$ mode of the fiber. 

Therefore, the SMF more efficiently uses existing the degrees of freedom provided by the DM to suppress a speckle, resulting in a significant improvement in starlight rejection over traditional speckle nulling using an imaging camera. For instance, a non-zero incident electric field that is antisymmetric about the center of the fiber tip will be eliminated in the overlap integral (Eq.~\ref{eq2}). We find that our optimization procedure often converges to solution where a node in the field or phase singularity appears at the fiber location, which is reminiscent of fiber nulling concept presented in \citet{Haguenauer2006}.

Generally speaking, a single-mode fiber will couple less starlight on average than a multi-mode fiber or detector resolution element (resel) with equivalent planet coupling/detection capability. Mathematically, neighboring speckles in a bandlimited complex stellar field have opposite parity, and therefore the overlap integral between the fundamental mode of the SMF and the stellar electric field is always less than or equal to the equivalent stellar energy on a resel at the same location when the resel size is chosen to collect the same planet signal as the SMF. 

In other words, the nulling condition over the resel requires the anti-speckle field $A_{as}(x,y)$ to be exactly opposite to the speckle field $A_{s}(x,y)$, i.e.~$A_{as}(x,y)=-A_{s}(x,y)$ for all spatial position over the resel $(x,y)$. The nulling condition through the single-mode fiber from Eq.~\ref{eq2}, implies that the resulting complex field  $A_{as}(x,y)+A_{s}(x,y)=A(x,y)$ is zeroed when projected onto and integrated over the single-mode fiber fundamental mode $\mathrm{HE}_{11}(x,y)$, which is a much less stringent and thus easier condition to meet.

\subsubsection{Throughput losses after nulling}
\label{subsubsection:Throughput_losses_after_nulling}
Referring to Eq.~\ref{eq1}, one must be careful that the planet signal does not suffer throughput losses in the starlight suppression process. To improve the performance of our HDC system, we must achieve a starlight suppression ratio that is better than the square of the planet throughput loss ratio. It is therefore significant that we have routinely demonstrated a factor of $>100$ suppression of starlight with no detected loss of planet throughput, effectively improving the SNR by a factor of $>10$.

From Eq.~\ref{eq2}, the throughput of the planet signal is roughly proportional to the Strehl ratio of the planet PSF described by the field $A(x,y)$. Using the Marechal approximation \citep{Mahajan1982}, we have
\begin{equation}\label{eq4}
    \eta\propto \exp{-\left(\frac{2\pi}{\lambda}\frac{h_0}{2\sqrt{2}}\right)^2},
\end{equation}
where $h_0/2$ is the amplitude of sinusoid we impose on the DM ($h_0$ on the wavefront). To constrain losses to be $< 10\%$, the ripple amplitudes are limited to $h_0 < \lambda/7\approx90$ nm during the speckle nulling process. By imposing sinusoidal shapes on the DM, in particular at low spatial frequencies (small angles), we might introduce slight displacements in the planet and star beams, which could easily translate into optical aberrations (e.g., when traveling through lenses on the bench). In extreme cases, this could degrade the beam quality and may even cause lateral beam shifts, which would effect the coupling efficiency and hence the throughput of the planet signal, as SMF coupling is sensitive to beam quality, angular offsets, and translational deviation. 

\subsubsection{Projected broadband performance}

\begin{figure*}[!t]
  \centering
\includegraphics[height=0.27\linewidth]{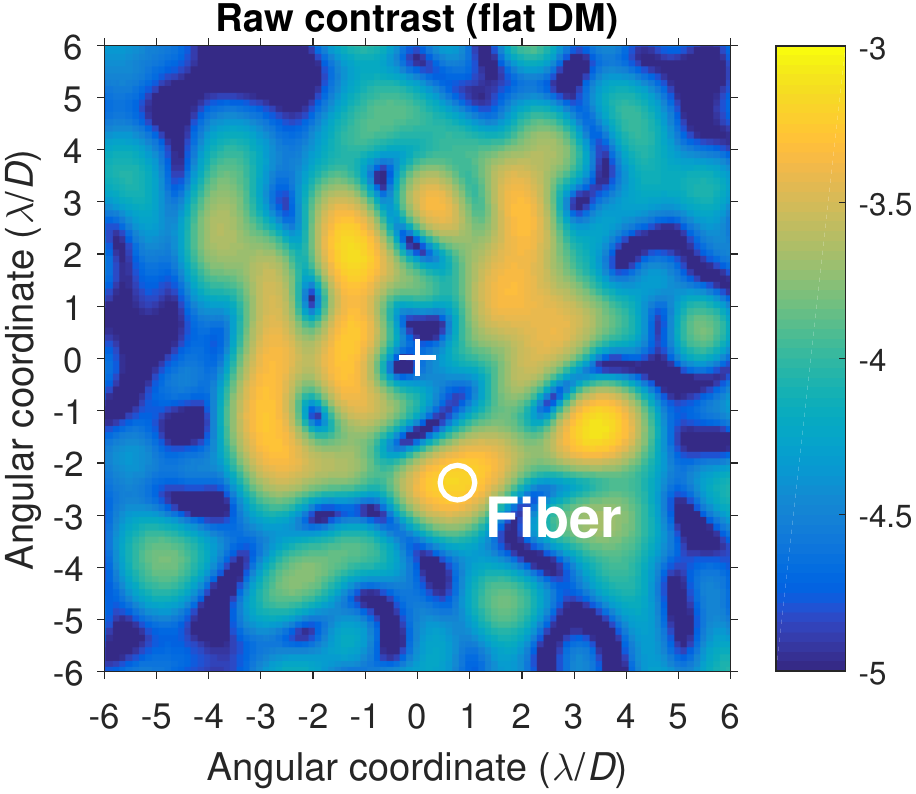}\hspace{2mm}
\includegraphics[height=0.27\linewidth]{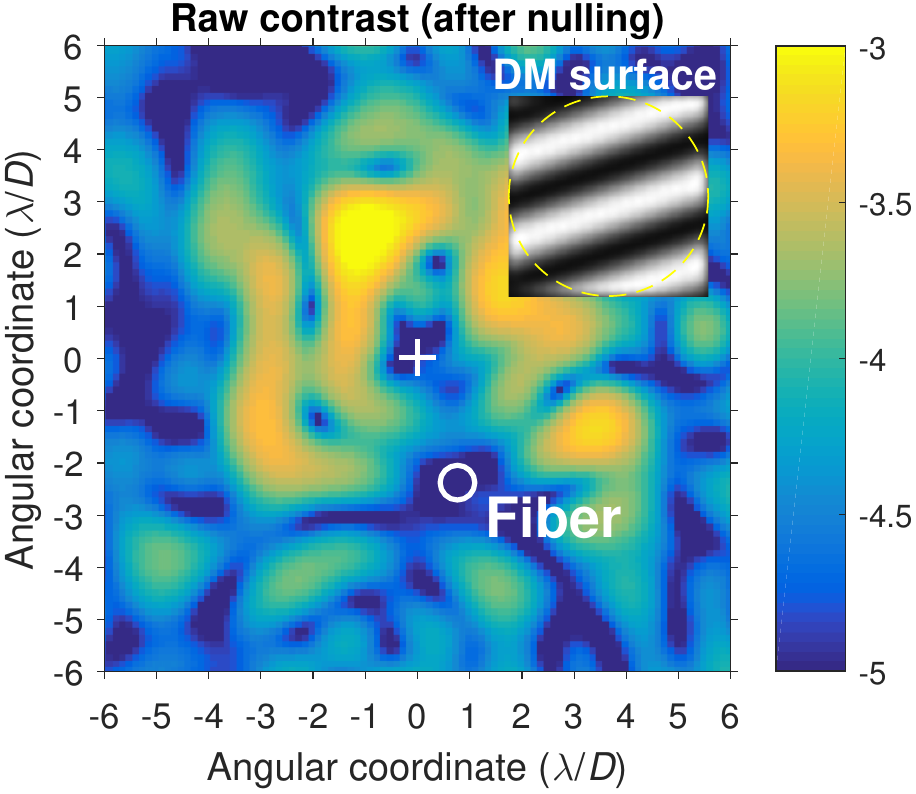}\hspace{2mm}
\includegraphics[height=0.27\linewidth]{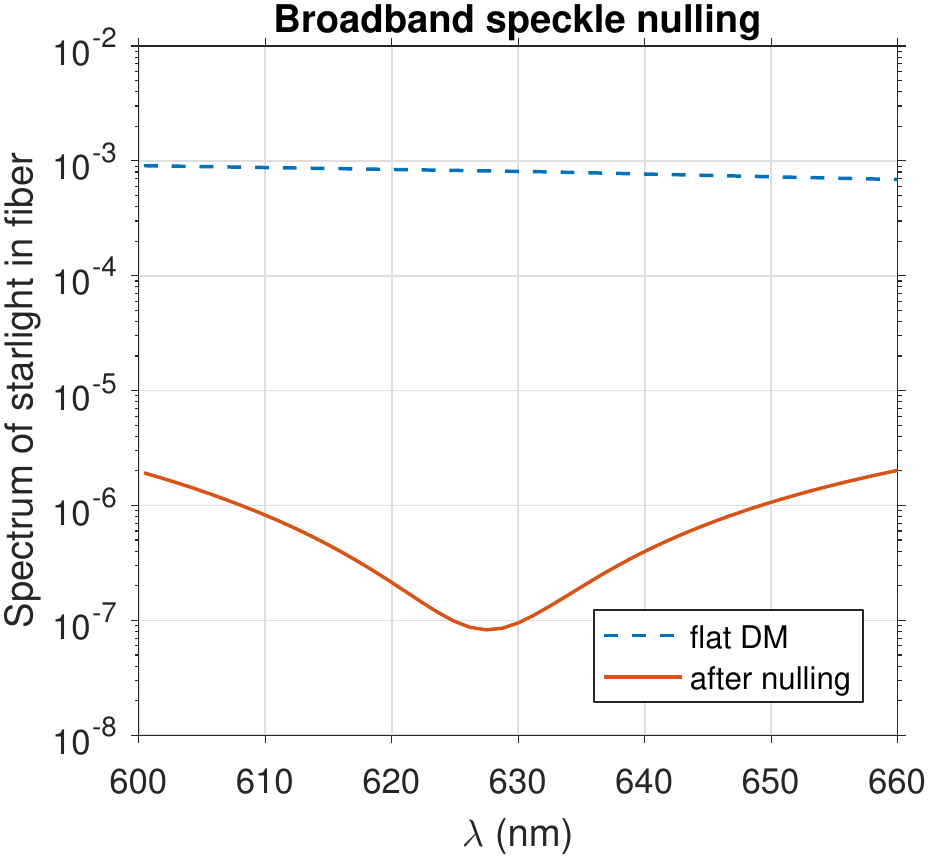}
  \caption{Simulation results in polychromatic light. Left:~Residual starlight after the coronagraph, with a flat DM surface. Middle:~Residual starlight after speckle nulling at the fiber location (2.5$\lambda/D$). The optimal DM surface is shown in the inset along with the footprint of the beam (yellow circle). Right: Spectrum of the residual stellar power coupled into the SMF before (flat DM) and after speckle nulling, normalized to the total stellar power before the coronagraph. \label{fig:broadband_sim}}
\end{figure*}

To confirm that this principle is readily extended to polychromatic light, we performed a series of numerical simulations that mimic our experimental setup, but allow us to explore the performance with varying source properties, fiber positions, and levels of optical aberration. Figure \ref{fig:broadband_sim} shows the simulated stellar PSF before and after the speckle nulling process, as well as the measured spectrum in the spectrograph, over a 10\% passband centered at 632~nm. The simulated optical system applied $\sim$30~nm~rms wavefront error to the beam. The fiber was placed at the location of a bright speckle and the coupled power was minimized at five discrete wavelengths. In doing so, a suppression factor of $>$100 is achieved across the full passband (see Fig. \ref{fig:broadband_sim}, right). We also verified that the planet throughput did not decrease due to the speckle correction. In fact, the planet throughput in this particular case increased by 0.5\%.

\begin{figure*}[!t]
  \centering
\includegraphics[trim={1.1cm 0.5cm 1cm 1cm},clip,width=1.025\linewidth]{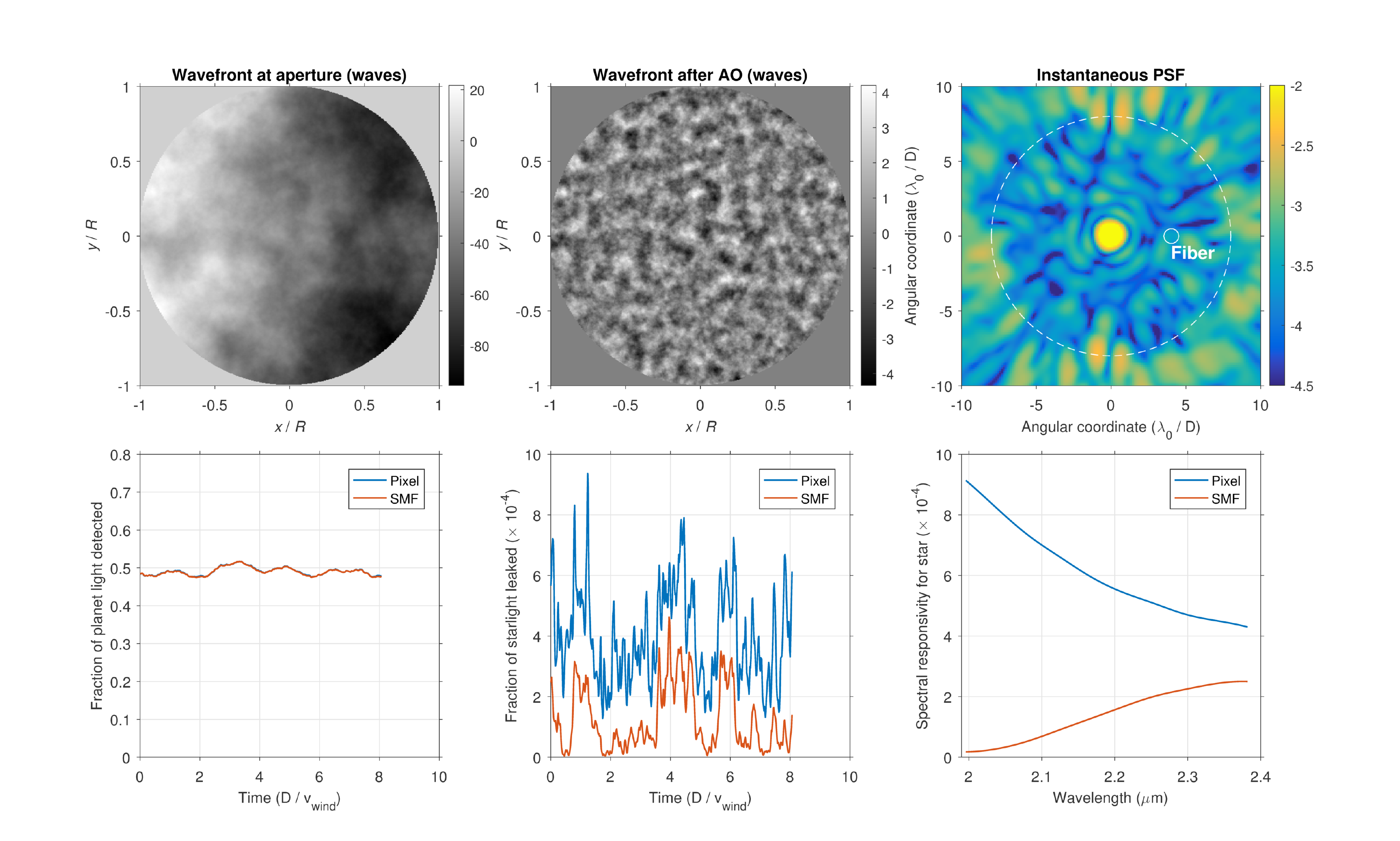}
  \caption{(Movie online) Numerical simulation of stellar leakage into a SMF, compared to an equivalent detector pixel, after an AO system. Top row: (left) The wavefront at the telescope pupil, given by a Kolmogorov phase screen. (middle) The wavefront after AO with 250~nm rms error. (right) The instantaneous stellar PSF with the control region and fiber location indicated by white circles. Bottom row: (left) The fraction of planet light coupled into the SMF and sensed by a single pixel. The size of the pixel was chosen such that these quantities are approximately equal. Time is shown in units of the clearing time $D/v_\mathrm{wind}$ under the frozen flow approximation, where $D$ is the telescope diameter and $v_\mathrm{wind}$ is the wind speed. (middle) The fraction of starlight coupled into the SMF and sensed by a single pixel. On average, 3$\times$ more starlight is sensed by a detector pixel than coupled into a SMF. (right) The instantaneous spectral responsivity for the star. \label{fig:SMFbehindAO}}
\end{figure*}

\subsubsection{Projected on-sky performance with a passive FIU}

 We performed a simulation where the amount of starlight passively sensed by a SMF and an equivalent single pixel was monitored over time in the presence of atmospheric turbulence (see Fig. \ref{fig:SMFbehindAO}). The effect of the AO was modeled by high pass filtering a Kolmogorov phase screen up to 8 cycles per pupil diameter $D$ (equivalent to a deformable mirror with 16 actuators) giving a post-AO wavefront error of $\sim$250~nm rms. Time lag and atmospheric chromaticity are ignored and should not affect the outcome of the numerical experiment. The fiber and pixel were placed at the same arbitrary location within the AO control region. We found that over 10 clearing times under the frozen flow approximation ($D/v_\mathrm{wind}$, where $v_\mathrm{wind}$ is the wind speed, typically $\sim$1~sec), an average of 3$\times$ less starlight was coupled into the SMF than sensed by the single pixel, without additional speckle nulling.

\subsubsection{Projected on-sky performance with an active FIU}
We conjecture that implementing a speckle nulling procedure with temporal bandwidth $>v_\mathrm{wind}/D$ and predictive control \citep{Poyneer2007, Riggs2016} may provide an additional factor of 10-100$\times$ in starlight suppression (Guyon \& Males 2017, submitted to ApJ), taking advantage of the natural speckle rejection provided by the SMF. We defer further analysis of the active FIU in the presence of dynamical aberrations to a forthcoming paper.

\section{Perspectives}
The fiber injection unit described here will be the core of the Keck Planet Imager and Characterizer instrument \citep[KPIC,][]{Mawet2016}. KPIC is a four-pronged upgrade of the Keck adaptive optics facility. The first stage is the addition of a high performance small inner working angle L-band vortex coronagraph to NIRC2 \citep{Absil2016,Mawet2017,Serabyn2017}, implemented in 2015 and now available to the Keck community in shared-risk mode. This upgrade not only included a brand new coronagraph mask, but also a suite of software packages to entirely script the coronagraph acquisition procedure, including automatic ultra-precise centering \citep{Huby2015}, speckle nulling wavefront control \citep{Bottom2016a}, and an open source python-based data reduction package \citep{Gomez2016}. 

The second upgrade component is an infrared pyramid wavefront sensor demonstration and potential facility for the Keck II adaptive optics system. Near-infrared wavefront sensing is a critical technology for science with AO on current and future telescopes. It will enable high-contrast observations of exoplanets around low-mass stars and in obscured starforming regions. It can be used to extend the performance of natural guide star (NGS) AO to redder targets and to increase the sky coverage of laser guide star (LGS) AO. Furthermore, it allows the application of optimal wavefront sensing approaches (e.g. pyramid and Zernike wavefront sensing) due to the AO correction at near-infrared wavelengths. This demonstration is especially relevant because all of the extremely large telescopes (ELTs) are planning to use infrared wavefront sensing as part of their AO facilities. 

The third upgrade is a higher-order deformable mirror paired with the infrared pyramid sensor, followed by a new single-stage coronagraph. The vortex coronagraph installed with the first upgrade component is inside of the NIRC2 cryostat and thus cannot be used in conjunction with a high-resolution spectrograph. Finally, the fourth component of the KPIC is the fiber injection unit discussed in this paper. The second, third, and fourth module will be integrated within the same optical relay.

KPIC is thus a phased, cost-effective upgrade path for the Keck II adaptive optics facility, building on the lessons learned from first- and second-generation high-contrast adaptive optics instruments, meant to explore new scientifically exciting niches and pave the way for the TMT-Planet Finder Instrument (PFI) core science, while maturing system-level and critical components for future ground- and space-based instrumentation, including NASA's HabEx and LUVOIR flagship mission concepts.

\subsection{Characterization of known objects}

The FIU concept presented here is amenable to the characterization of exoplanets discovered by other methods (direct imaging, RV, astrometry, etc.) where the  planet position is known a priori. Our proposed pointing and tracking system is accurate enough to offset blindly to the location of a companion too faint to be visible in acquisition images. To further improve pointing astrometric accuracy, one could use the deformable mirror to generate a set of satellite spots as routinely used by VLT/SPHERE or Subaru/SCExAO \citep{Jovanovic2015b}. 

\subsection{Multiplexing}

The HDC technique may also be multiplexed to increase the effective field of view, and so it can be used to detect new planets, or characterize planets whose positions are not well constrained, such as the radial velocity detected Earth-like planet around Proxima Centauri \citep{Lovis2016}. One limitation for spatial multiplexing is detector real estate. Preliminary design work has led us to consider a 3x3, 9-element multiplexing capability using a H4RG detector (4096x4096 pixels). The sampling in the image plane is done by way of a microlens array as in \citet{Ireland2014,Rains2016}, where each microlens feeds a single mode fiber. The fiber output may be reconfigured in a pseudo-slit at the entrance of an echelle spectrograph. \citet{Rains2016} recently demonstrated a 3x3 lenslet-based 9-single mode fiber mini integral field unit linking Subaru/SCExAO to the RHEA spectrograph \citep{Bento2016}. However, this first attempt was affected by modal noise and cross talk due to the small spacing between fibers in the output pseudo-slit.

Another multiplexing option currently proposed is to build as many high-resolution diffraction-limited spectrographs as there are resolution elements in the search area field of view. Diffraction-limited spectroscopy is economical due to the conservation of beam etendue and is likely to be the most realistic implementation of future high-resolution spectrographs on large telescopes \citep{Bland-Hawthorn2004,Bland-Hawthorn2006,Bento2016}.

We note that speckle nulling on multiple single-mode fibers should still work. However, the number of available degrees of freedom per fiber will be smaller, and likely result in reduced starlight suppression gains.

\section{Conclusion}

In this paper, we presented an innovative fiber injection unit module designed to efficiently couple a high-contrast imaging system (adaptive optics and coronagraph) to a high-resolution spectrograph, enabling high-dispersion coronagraphy (HDC) of exoplanets. We built a first FIU prototype and performed a series of laboratory experiments that demonstrated fast off-axis planet light acquisition, as well as high ($\simeq 70\%$) and stable coupling efficiencies. Using the wavefront control system and a technique akin to speckle nulling, we achieved high levels of starlight suppression. Using the extreme modal selectivity of single-mode fibers, we routinely obtained speckle suppression gains that outperform conventional image-based speckle nulling by at least two orders of magnitude. Our FIU demonstrator is a prelude to on-sky scientific demonstrations with the KPIC project, a pathfinder to future HDC instruments on extremely large telescopes on the ground and in space.

\acknowledgments

The authors would like to acknowledge the referee for his thorough review and constructive comments. The authors would like to acknowledge the financial support of the Heising-Simons foundation.

\bibliography{fiu_arxiv_bib}


%% in parentheses, after the keyword, but they will not be verified.

{\it Facilities:} \facility{Caltech's High-Contrast Spectroscopy Testbed for Segmented Telescopes (HCST).}



\end{document}